\documentstyle[aps,floats,epsf,twocolumn]{revtex}
\begin{document} \draft

\title{Self-consistent electronic structure of a $d_{x^2-y^2}$ 
and a $d_{x^2-y^2}+id_{xy}$ vortex}

\author{M. Franz and Z. Te\v{s}anovi\'c}
\address{Department of Physics and Astronomy, Johns Hopkins University,
Baltimore, MD 21218
\\ {\rm(\today)}
}
%
\address{~
\parbox{14cm}{\rm
\medskip
We investigate quasiparticle states associated with an isolated 
vortex in a $d$-wave superconductor 
using  a self-consistent Bogoliubov-de Gennes formalism. 
For a pure $d_{x^2-y^2}$ superconductor we find that 
there exist no bound states in the core; 
all the states are extended with continuous energy spectrum. This result is
inconsistent with the existing experimental data on cuprates. 
We propose an explanation for this data in terms of a 
magnetic-field-induced  
$d_{x^2-y^2}+id_{xy}$ state recently invoked in connection with the thermal
conductivity measurements on Bi$_2$Sr$_2$CaCu$_2$O$_8$.
}}
\maketitle


%
\narrowtext

Behavior of superconductors in a magnetic field has been traditionally at the
center of  the condensed matter research because of
the rich variety of fascinating phenomena in such systems and also
because of its considerable technological importance. With the advent of
high-$T_c$ cuprates the subject became even more interesting owing to the  
realization that their order parameter is unconventional, exhibiting
most likely  a $d_{x^2-y^2}$ symmetry. A number of novel effects 
associated with the existence of low-lying quasiparticle excitations near 
the gap nodes  have been predicted\cite{yip,volovik,berlinsky}.
While there exists experimental evidence for each of these 
effects\cite{maeda,moler,fischer}, it is by no means conclusive and an
active debate continues. Several fundamental issues remain unresolved.
Among the most interesting is the electronic structure of 
magnetic vortices, where a coherent physical picture is still lacking 
despite the number of theoretical and experimental investigations. 
Notably the nature of quasiparticle core states in
$d$-wave superconductors remains
poorly understood. These states are of considerable
importance as they impact on various static and dynamic
properties of the mixed state such as the 
 structure of the flux lattice, pinning and flux-flow resistance.  

It was established more than three decades ago
by Caroli, de Gennes and Matricon\cite{caroli} that discrete 
quasiparticle states exist localized in vortex cores 
of conventional $s$-wave superconductors. 
These states are labeled by the angular momentum quantum 
number $\mu$, and the low-lying eigenvalues are
$E_\mu\simeq \mu(\Delta_0^2/E_F)$ with $\mu=1/2,3/2,\dots$, $\Delta_0$ the
bulk gap and $E_F$ the Fermi energy.  This prediction has been later 
confirmed in  
detail by numerical computations\cite{gygi} and by experiments on
NbSe$_2$ \cite{hess}. 
 In a $d_{x^2-y^2}$ superconductor 
the situation becomes considerably more complex owing to the non-trivial
structure of the gap function which vanishes along the four nodes on
the Fermi surface. While the 
$s$-wave bound states can be intuitively understood 
by drawing an analogy to a simple quantum-mechanical problem of a particle
in the cylindrical well of the radius $\xi\simeq v_F/\pi\Delta_0$ and height
$\Delta_0$, a suitable analogy in a $d$-wave
case would involve a potential well whose radius and height
depend on the polar angle $\theta$, with $\Delta(\theta)$ vanishing along the
four diagonal directions causing $\xi(\theta)$ to diverge. 
Under such circumstances
one would expect low-lying states to be {\em extended}
along the node directions, rather than localized, unless the mixing
of the core levels into the continuum is prevented by some higher symmetry. 
The existing experimental work, however,  appears consistent with 
{\em localized} core states with large energy 
spacing\cite{fischer,karrai,jiang,note1}. 

The problem of a vortex in a $d_{x^2-y^2}$ superconductor 
has been considered theoretically by a number of authors
\cite{soininen,wang,maki1,ichioka,kopnin} 
but the detailed nature of quasiparticle core states has not been addressed. 
Very recently Morita, Kohmoto and Maki\cite{mkm}  studied this problem
using an approximate version of the Bogoliubov-de Gennes (BdG)
 theory. As pointed out in
a subsequent Comment\cite{franz}, this approximation improperly 
neglects an essential element of the physics of $d_{x^2-y^2}$ superconductors
and yields unphysical results.

In this article we present, for the first time, 
 a fully self-consistent numerical solution of
the continuum  BdG theory for a single isolated
 vortex in a $d$-wave superconductor.
In agreement with the above qualitative argument we find that for a pure
$d_{x^2-y^2}$ gap there exist no {\em truly localized} core 
states\cite{note3}. We find low lying
states that are strongly peaked in the core but have tails along the gap node
directions which do not appear to decay to zero far from the core.  
The energy spectrum of these states becomes continuous in the limit 
of infinite system size. This is consistent with
their extended nature, but inconsistent with the experimental finding
\cite{fischer,karrai} of a large gap, $E_0\approx\Delta_0/5$, to the lowest
core state in YBa$_2$Cu$_3$O$_{7-\delta}$. 
We propose that one can reconcile the theory with experiment
by assuming a parity and time reversal symmetry violating 
$d_{x^2-y^2}+id_{xy}$ state which has been recently invoked to explain 
the thermal conductivity data on Bi$_2$Sr$_2$CaCu$_2$O$_8$ in finite
magnetic fields
\cite{krishana,laughlin}. Such a state is fully gapped and for sufficiently 
large $d_{xy}$ component it can support true bound states in the core
of the nature similar to the $s$-wave case. 

The BdG equations for a $d$-wave vortex have been previously solved numerically
on the lattice\cite{soininen,wang}. In this work we adopt a continuum
formulation of the problem, which has been used to study conventional 
$s$-wave vortices\cite{gygi}. Such formulation allows us to directly exploit
various symmetries of the system and as a result we are able to 
consider much larger systems with a reasonable spectral resolution.
The BdG equations for a $d$-wave superconductor can be written 
as\cite{degennes}
\begin{eqnarray}
\hat{\cal H}_e u({\bf R}) +\int d{\bf r}\Delta({\bf R}-{\bf r}/2,{\bf r})v({\bf R}-{\bf r})&=& Eu({\bf R}),
\nonumber\\
-\hat{\cal H}^*_e v({\bf R}) +\int d{\bf r}\Delta^*({\bf R}-{\bf r}/2,{\bf r})u({\bf R}-{\bf r})&=& Ev({\bf R}).
\label{bdg}
\end{eqnarray}
Here $\hat{\cal H}_e$ is the single electron Hamiltonian, which we assume 
to have a simple free particle form 
\begin{equation}
\hat{\cal H}_e={1\over 2m}\left({\bf p}-{e\over c}{\bf A}\right)^2-E_F,
\label{he}
\end{equation}
and $\Delta({\bf R},{\bf r})=V({\bf r})\langle c_\uparrow({\bf R}-{\bf r}/2)
c_\downarrow({\bf R}+{\bf r}/2)\rangle$ is 
the order parameter which is a function of the center-of-mass and the relative
coordinate ${\bf R}$ and ${\bf r}$. In terms of the 
quasiparticle amplitudes $[u_n,v_n]$
the order parameter is self-consistently determined from the gap equation
\begin{eqnarray}
\Delta({\bf R},{\bf r})&=&V({\bf r}){\sum_n}'[u_n({\bf R}-{\bf r}/2)v_n^*({\bf R}+{\bf r}/2)\nonumber \\
&+&u_n({\bf R}+{\bf r}/2)v_n^*({\bf R}-{\bf r}/2)]\tanh(E_n/2T),
\label{gap}
\end{eqnarray}
where $n$ labels the eigenstates of (\ref{bdg}) and the prime means the usual
restriction to energies smaller than a cutoff scale $\Omega_c$. 
$V({\bf r})$ is the pairing interaction which
we assume to have the following model form:
\begin{equation}
V({\bf r})=V_0\delta({\bf r})+V_1g(\varphi){1\over a}\delta(r-a),
\label{int}
\end{equation}
with ${\bf r}=(r,\varphi)$. The potential (\ref{int})
consists of a contact repulsion term $V_0>0$, meant to model the on-site
Coulomb repulsion between holes, and a short range attractive part $V_1<0$,
necessary to establish superconducting pairing. In order to model the lattice
structure of cuprates we allow this attraction to be angle dependent with 
\begin{equation}
g(\varphi)=(1-\epsilon)\cos^2(2\varphi) + \epsilon\sin^2(2\varphi),
\label{ang}
\end{equation}
where $\epsilon$ is a parameter used to tune the relative strength of this
model's analog of nearest and second nearest neighbor attraction. 
The nature of superconducting instability in this model depends on 
the dimensionless parameter $\eta= ak_F$. In the absence 
of magnetic field the ground state has a $d$-wave symmetry 
for $\eta>1.9$ and is extended $s$-wave for $\eta<1.9$. This is consistent 
with the phase diagram of the related lattice model, where $d$-wave is
known to be stable only close to half filling\cite{feder}. Furthermore, in the
$d$-wave regime, the ground state is a pure $d_{x^2-y^2}$ for $\epsilon=0$, 
a pure $d_{xy}$ for $\epsilon=1$, and a  $d_{x^2-y^2}+id_{xy}$ admixture
for $0<\epsilon<1$. More generally one can also allow for anisotropy 
in the single particle Hamiltonian $\hat{\cal H}_e$ to account for 
the band structure effects in cuprates. We find that such anisotropies 
do not modify the phase diagram significantly and have only secondary effect
on the vortex core
structure. In the following we therefore limit ourselves to the 
simple form (\ref{he}) and we furthermore
 neglect the vector potential ${\bf A}$,
as appropriate for the extreme type-II cuprates\cite{caroli}.
The magnitude of the contact repulsion $V_0$
has little effect on the physics in the $d$-wave regime\cite{feder}.
 For simplicity we therefore take $V_0=0$. 

The form of the pairing
interaction (\ref{int}) implies that the order parameter depends on the
relative coordinate ${\bf r}$ only through its polar angle $\varphi$. It is 
convenient to expand it in terms of the 2D angular momentum eigenstates,
\begin{equation}
\Delta({\bf R},{\bf r})=\Delta(R,\theta;\varphi)=
\sum_{p,l}e^{-ip\theta}e^{il\varphi}\Delta_{pl}(R),
\label{gape}
\end{equation}
where ${\bf R}=(R,\theta)$. In such a  
representation the integer $p$ characterizes winding of the superconducting 
phase around the vortex and  $l$ specifies the orbital state of the 
Cooper pair. Thus, for instance, the dominant order parameter
near a singly quantized $d_{x^2-y^2}$ vortex  consists of an equal 
superposition of $p=1$, $l=\pm 2$ components: $\Delta(R,\theta;\varphi)=
2e^{-i\theta}\cos(2\varphi)\Delta_{1,\pm2}(R)$.

We solve the BdG equations (\ref{bdg}) numerically on a disk of the radius
$R_0$ with a suitably chosen initial order parameter and we then iterate
Eqs.\ (\ref{bdg}) and (\ref{gap}) until self-consistency is achieved.
Following Gygi and Schluter\cite{gygi}
we expand the quasiparticle amplitudes in the basis spanned by 
the eigenfunctions of $\hat{\cal H}_e$:
\begin{equation}
[u({\bf R}),v({\bf R})]=\sum_{\mu,m}e^{i\mu\theta}\Phi_{\mu m}(R)[u_{\mu m},v_{\mu m}].
\label{exp}
\end{equation}
Here $\Phi_{\mu m}(R)=[\sqrt{2}/
R_0J_{\mu+1}(\alpha_{\mu m})]J_{\mu}(\alpha_{\mu m}R/R_0)$, $J_{\mu}(z)$ is
the Bessel function of order $\mu$ and $\alpha_{\mu m}$ is the $m$-th zero
of $J_{\mu}(z)$. The integro-differential equation (\ref{bdg}) thus becomes 
an eigenvalue problem with an infinite matrix which 
we truncate at large values of $|\mu|$ and $m$, and diagonalize using 
a standard LAPACK subroutine. In the $s$-wave case this matrix is
block diagonal in the angular momentum $\mu$ and the resulting 
radial problem can be solved for each $\mu$ separately. 
The crucial new element in the
$d$-wave case is that, due to the complicated structure of the pairing term
(\ref{gape}),
the angular momentum channels remain coupled and one has to solve the 
{\em full} 2D problem. This essential feature was ignored in \cite{mkm}.

\begin{figure}[t]
\epsfxsize=8.5cm
\epsffile{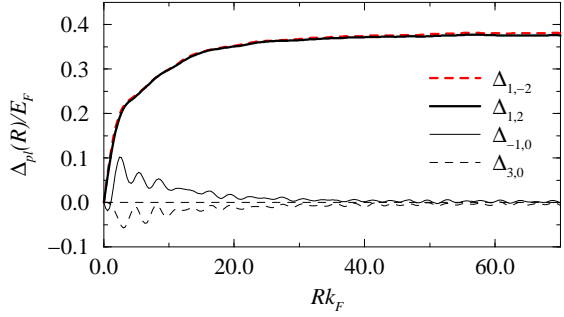}
\caption[]{Dominant order parameter components for system
size $R_0k_F=120$. In terms of conventional notation we have 
$|d_{x^2-y^2}|\propto(\Delta_{1,-2}+\Delta_{1,2})$,
$|d_{xy}|\propto(\Delta_{1,-2}-\Delta_{1,2})$ and
$|s|\propto|e^{i\varphi}\Delta_{-1,0}+e^{-3i\varphi}\Delta_{3,0}|$.
}
\label{fig1}
\end{figure}

To model a pure $d_{x^2-y^2}$ case we choose $\eta=2$, $\epsilon=0$, 
$\Omega_c/E_F=0.3$, $V_1/E_F=1.6$ and $T=0$.  
We assume the initial order parameter of the form
$\Delta_{1,\pm 2}(R)=\Delta_0\tanh(R/\xi_0)$,
where $\xi_0=v_F/\pi\Delta_d$ is the coherence length\cite{note2}. 
The above parameters imply $\Delta_d=0.26E_F$ and $\xi_0=2.5k_F^{-1}$. 
In the
process of iterating  Eqs.\ (\ref{bdg}) and (\ref{gap}) various other 
components $\Delta_{pl}(R)$ appear in the self consistent solution 
reflecting the spatial anisotropy of the
dominant $d_{x^2-y^2}$ component and nucleation of various sub-dominant
order parameters near the core. Fig. \ref{fig1} shows the leading 
components of the self-consistent order parameter. 
The $d_{x^2-y^2}$
order parameter relaxes to its bulk value over a distance $\sim20k_F^{-1}$,
which is much larger than $\xi_0$. In fact, by varying the coupling strength
$V_1$ in a wide range,
 we find that for coherence lengths $\xi_0k_F\lesssim 10$ 
the order parameter profile (rescaled to its bulk value)
no longer depends on $\xi_0$ and retains the universal shape 
showed in Fig. \ref{fig1}. We also find
that this profile is not very well fitted by the usual $\tanh(R/\xi')$ function
for any $\xi'$ but is more consistent with algebraic $\sim1/r^2$ relaxation. 
Fig. \ref{fig1} also shows sizable components with $s$-wave symmetry of the
form predicted by the Ginzburg-Landau theory\cite{berlinsky}. We have 
explicitly verified that, in agreement with \cite{berlinsky}, they will form
four satellite vortices at a distance 
$\sim26 k_F^{-1}$ form the origin. 

\begin{figure}
\epsfxsize=8.5cm
\epsffile{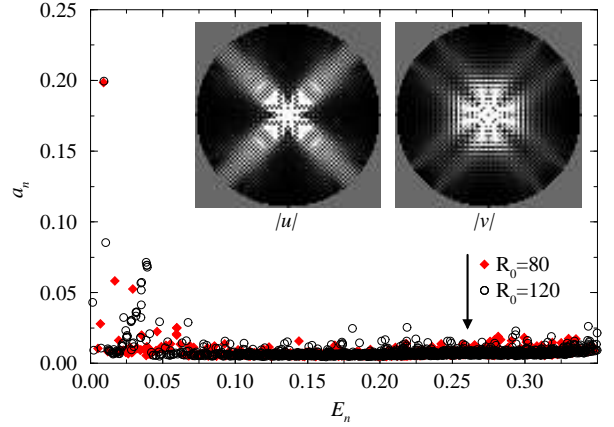}
\caption[]{Inverse participation ratios $a_n$ as a function of energy.
 The arrow marks the maximum bulk gap $\Delta_d$.
Inset: amplitudes $|u_n({\bf R})|$ and $|v_n({\bf R})|$ of the lowest core 
state. 
}
\label{fig2}
\end{figure}

We now investigate the nature of quasiparticle core states by computing 
the generalized inverse participation ratios, defined 
as\cite{franz2},
\begin{equation}
a_n=(R_0k_F)^2{\langle |u_n|^4\rangle_s + \langle |v_n|^4\rangle_s
     \over (\langle |u_n|^2\rangle_s + \langle |v_n|^2\rangle_s)^2},
\label{part}
\end{equation}
where $\langle \dots \rangle_s=\int \dots d{{\bf R}}$. 
As a function of increasing system size $R_0$ this quantity 
approaches a finite constant for an extended state and grows as $R_0^2$ for a 
state localized within the characteristic length $\xi_L\ll R_0$.
Fig.\ \ref{fig2} shows $a_n$ as a function of energy for system sizes
$R_0k_F=80,120$.
Over the entire energy range the data for two different sizes behave in a 
similar way. If there existed localized states in the core
their corresponding $a_n$ would have increased by more than a factor of $2$ 
between sizes 
$R_0k_F=80$ and $120$. No such increase is observed. 
We carried out similar analysis for an $s$-wave case (same 
parameters with $\eta=1$), where it is known 
that only localized states exist at low energies. Indeed,
in this case a clear $a_n\sim R_0^2$ scaling was observed\cite{mf} for 
$E_n < \Delta_0$. 
Fig.\ \ref{fig2} further shows that, unlike in the  $s$-wave case\cite{gygi},
there is no discernible pattern in the core energy levels and their spacing 
decreases with increasing system size. Visual inspection of the 
amplitudes of these states (inset Fig.\ \ref{fig2}) reveals that they are 
highly anisotropic with peaks near the core and 
tails running along the gap node directions which appear to saturate to a
{\em finite amplitude} far from the core.  
Similar states are known to exist near a strongly scattering non-magnetic
impurity\cite{salkola}.

\begin{figure}
\epsfxsize=8.5cm
\epsffile{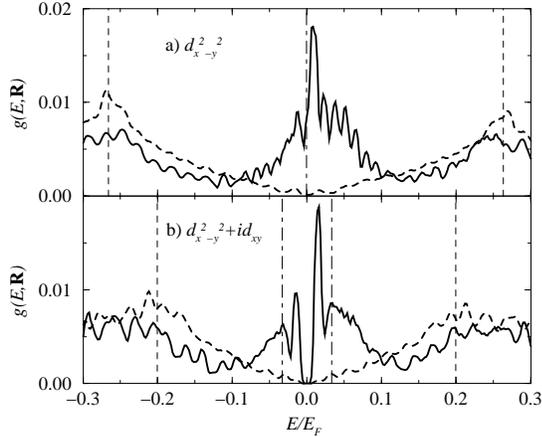}
\caption[]{Tunneling conductance $g(E,{\bf R})$ at the center of the vortex 
(solid line) and far from the vortex (dashed line), both spatially averaged
over a circular ring $2k_F^{-1}$ wide. Dashed and dash-dotted vertical lines
mark the maximum and minimum bulk gap respectively; $R_0k_F=120$.
}
\label{fig3}
\end{figure}

Although not truly localized these core states should still give rise to 
enhanced tunneling conductance from the cores. In 
Fig.\ \ref{fig3}a we display this quantity computed from the present model. 
The closely spaced core states give rise to a broad peak centered near 
the zero 
energy. The peaks corresponding to individual states can still be 
resolved, but by studying the size dependence of such spectra we conclude that
in the limit of infinite size these will form a continuum, similar to that
reported by Wang and MacDonald in their lattice calculation\cite{wang}.
While the fine
details of the spectrum depend on parameters of the model as well as on the
system size, its gross qualitative 
features, and in particular the vanishingly small gap to the lowest state, 
remain remarkably robust for a wide range of parameters considered. 
These spectra are, unfortunately, in disagreement with the 
experimental finding of a large $\sim\Delta_d/5$ gap to the lowest core
state\cite{fischer,karrai}. It is in principle 
 possible that the intragap spectral
features found experimentally are in fact resonances rather that true bound 
states. Even so, we find no indication of such resonant behavior in our
calculation. Ref.\ \cite{fischer} also reports that the core
states are spatially isotropic, in 
contradiction to strong anisotropies found by this work and by quasiclassical
computations\cite{maki1,ichioka}.

It would thus appear
that a model based on a pure $d_{x^2-y^2}$ order parameter is in all aspects
{\em inconsistent} with the published 
experimental data. We therefore propose an explanation in terms
of a mixed $d_{x^2-y^2}+id_{xy}$ state, which may be formed in cuprates
in finite fields at low temperatures\cite{krishana,laughlin}. 
Such a state is fully gapped and we therefore expect 
the quasiparticle core states to recover some of their localized 
$s$-wave character.  Within the present model we study such a scenario
by taking $\epsilon=0.25$ (with other parameters unchanged). This results in 
a $d_{x^2-y^2}+id_{xy}$ state with  $d_{xy}\simeq 0.17 d_{x^2-y^2}$,
roughly consistent with Laughlin's prediction\cite{laughlin}
 for $T=0$ and $B=6$T.
The order parameter distribution near the core remains qualitatively
similar to that reported in Fig.\ \ref{fig1}, but now with unequal magnitudes
of $\Delta_{1,2}$ and $\Delta_{1,-2}$. Carrying out the analysis of 
participation ratios we find that the 
core states below the minimum gap $d_{xy}$
are truly localized and nearly isotropic, in
agreement with experiment. Fig.\ \ref{fig3}b shows the corresponding 
tunneling conductance,
which is now qualitatively consistent with experiment in that there 
exists a finite gap to the lowest core state which is independent of the 
system size. This gap is however still too
small to account for the experimental result; we find that the 
agreement becomes better for larger $d_{xy}$ component.
 Note that within the present model
the $d_{xy}$ component is brought about by change in the pairing interaction,
rather than magnetic field and it 
would therefore be present even at zero field. 
Although we expect that the qualitative features of the core 
states are insensitive to the origin of $d_{xy}$, a more satisfactory
model would correctly describe 
the transition to the time reversal breaking state as
a function of field. This would require explicitly including the 
vector potential term ${\bf A}$ in the kinetic energy (\ref{he}); the work on
this is in progress.  

Our results firmly establish the absence of localized vortex core states in
superconductors with a pure $d_{x^2-y^2}$ gap. We speculate that  
a sizable admixture of a subdominant order parameter
is needed to reconcile theory with experiment. A magnetic-field-induced 
$d_{xy}$ component appears to be the most acceptable choice at present. This
is by no means free of problems. A sufficiently large  $d_{xy}$ component 
could be difficult to justify since on general grounds one would expect
$d_{xy}/d_{x^2-y^2}$ to scale as the ratio of the corresponding critical 
temperatures\cite{laughlin}. Furthermore, a transition to the fully gapped 
$d_{x^2-y^2}+id_{xy}$ state  would  be observable in the specific heat or 
muon-spin-rotation measurements of the penetration depth, but no such 
effect has been reported\cite{moler,sonier}. In the present context of the 
vortex core states, however, 
our proposition can be easily tested by measuring the core spectra at 
lower fields (or higher temperatures). If the present spectra are indeed
characteristic of a $d_{x^2-y^2}+id_{xy}$ state a dramatic
change should be observable at the transition to the pure $d_{x^2-y^2}$ state.
	
The authors are indebted to A. V. Balatsky, 
A. J. Berlinsky and A. J. Millis for
helpful discussions.  This research was supported by NSF
grant DMR-9415549.


\begin{references}
\bibitem{note2} Here $\Delta_d$ is defined as a maximum gap on the Fermi 
surface which is related to the real space bulk order parameter $\Delta_0$
by $\Delta_d=2J_2(\eta)\Delta_0\simeq 0.7\Delta_0$ for $\eta=2$.
\end{references}
\end{document}